\documentclass{sigchi-ext}
\usepackage[T1]{fontenc}
\usepackage{textcomp}
\usepackage[scaled=.92]{helvet} % for proper fonts
\usepackage{graphicx} % for EPS use the graphics package instead
\usepackage{balance}  % for useful for balancing the last columns
\usepackage{booktabs} % for pretty table rules
\usepackage{ccicons}  % for Creative Commons citation icons
\usepackage{ragged2e} % for tighter hyphenation

\def\plaintitle{Visualization Tool for Environmental Sensing and Public Health Data}

\def\emptyauthor{}
\def\plainkeywords{Citizen science; visualization; air quality; public health}

\title{\plaintitle}

\numberofauthors{6}
\author{%
    \textbf{Yen-Chia Hsu, Jennifer Cross, Paul Dille, Illah Nourbakhsh}\\
    \affaddr{The Robotics Institute} \\
    \affaddr{Carnegie Mellon University} \\
    \affaddr{Pittsburgh, Pennsylvania, U.S.A.} \\
    \email{\{yenchiah, jcross1, pdille, illah\}@andrew.cmu.edu} 
	\vfil
    \textbf{Leann Leiter, Ryan Grode}\\
    \affaddr{SWPA Environmental Health Project}\\
    \affaddr{McMurray, Pennsylvania, U.S.A.}\\
    \email{\{lleiter, rgrode\}@environmentalhealthproject.org}
    \vfil}
    
\definecolor{linkColor}{RGB}{6,125,233}
\hypersetup{%
  pdftitle={\plaintitle},
  pdfauthor={\emptyauthor},
  pdfkeywords={\plainkeywords},
  bookmarksnumbered,
  pdfstartview={FitH},
  colorlinks,
  citecolor=black,
  filecolor=black,
  linkcolor=black,
  urlcolor=linkColor,
  breaklinks=true,
}

\begin{document}

%% For the camera ready, use the commands provided by the ACM in the Permission Release Form.
%\CopyrightYear{2007}
%\setcopyright{rightsretained}
%\conferenceinfo{WOODSTOCK}{'97 El Paso, Texas USA}
%\isbn{0-12345-67-8/90/01}
%\doi{http://dx.doi.org/10.1145/2858036.2858119}
%% Then override the default copyright message with the \acmcopyright command.
\copyrightinfo{\acmcopyright}
\CopyrightYear{2018} 
\setcopyright{rightsretained} 
\conferenceinfo{DIS'18 Companion}{June 9--13, 2018, , Hong Kong}
\isbn{978-1-4503-5631-2/18/06}
\doi{https://doi.org/10.1145/3197391.3205419}

\maketitle

% Uncomment to disable hyphenation (not recommended)
% https://twitter.com/anjirokhan/status/546046683331973120
\RaggedRight{} 

% Do not change the page size or page settings.
\begin{abstract}
To assist residents affected by oil and gas development, public health professionals in a non-profit organization have collected community data, including symptoms, air quality, and personal stories. However, the organization was unable to aggregate and visualize these data computationally. We present the Environmental Health Channel, an interactive web-based tool for visualizing environmental sensing and public health data. This tool enables discussing and disseminating scientific evidence to reveal local environmental and health impacts of industrial activities.
\end{abstract}

\keywords{\plainkeywords}

\category{H.5.m.}{Information Interfaces and Presentation (e.g. HCI)}{Miscellaneous}

\section{Introduction and Related Work}

\begin{figure}[t]
	\centering
	\includegraphics[width=1\columnwidth]{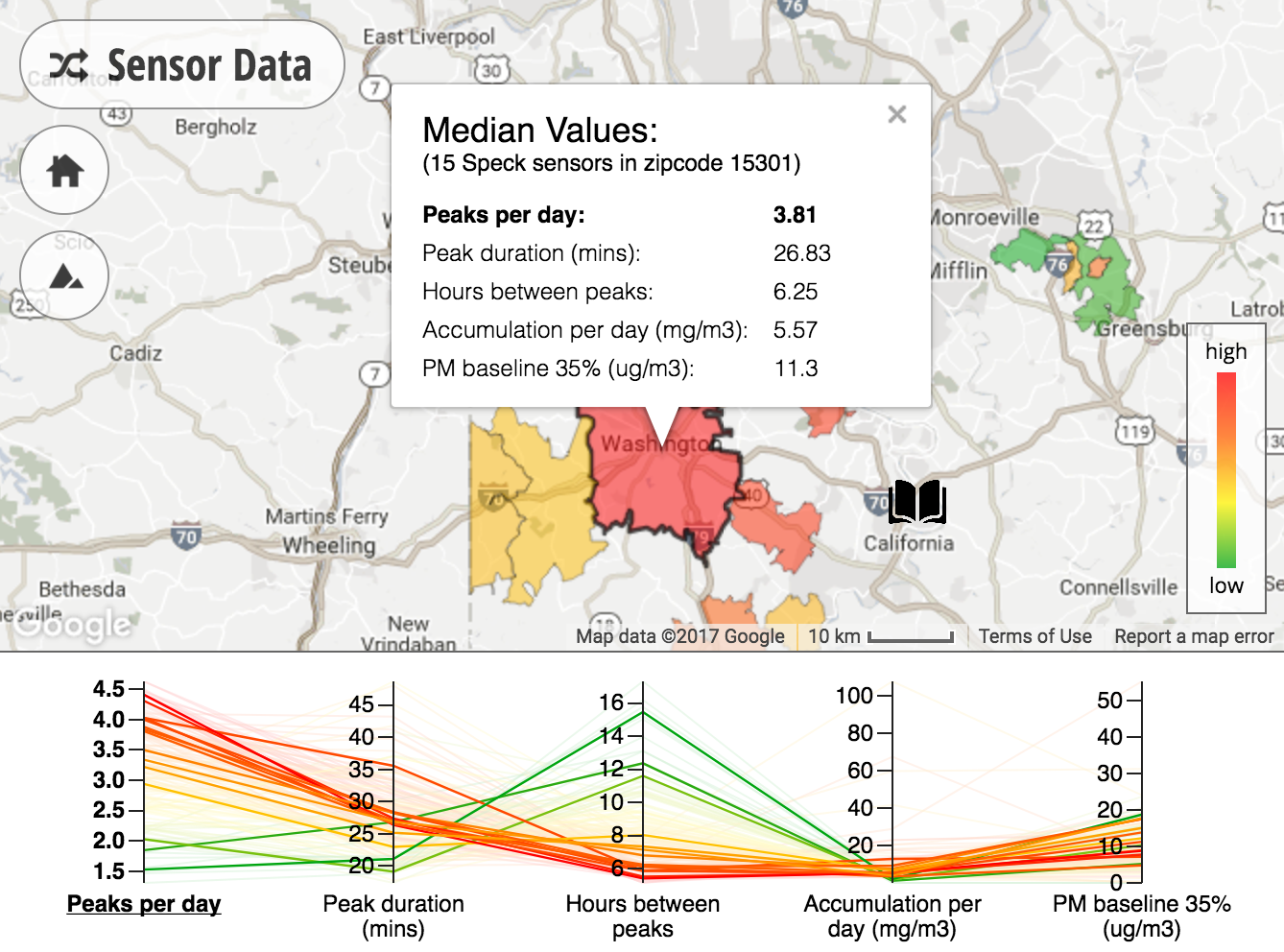}
	\vspace{-6mm}
	\caption{The user interface of the Environmental Health Channel, which visualizes the analysis of air quality sensors.}~\label{fig:UI}
	\vspace{-8mm}
\end{figure}

Air quality and its impacts on public health are critical environmental issues for residents who live near oil and gas development sites \cite{Colborn-2011}. A vital step towards addressing these issues is through the collection and dissemination of convincing scientific evidence of these impacts \cite{Hsu-2017, Hsu-2016}. However, conveying this evidence, especially with multiple types of data at a large temporal and geographic scale, requires the assistance of computational tools. In the pursuit of developing a tool for this purpose, we collaborated with a local non-profit organization that is working to study and assist communities that are potentially affected by oil and gas development. Since 2014, the organization has collected data which includes (1) particulate measurements from air quality sensors, (2) physical and psychosocial symptoms from surveys, and (3) personal stories from interviews. These citizen-contributed data were stored across multiple incompatible systems, which hindered retrieving information, visualizing trends, and disseminating findings. Moreover, the organization lacked the resources to independently develop computational tools for aggregating and visualizing data to facilitate user decision-making. Therefore, we collaborated with health professionals from the non-profit organization to develop the Environmental Health Channel (EHC), an interactive web-based data visualization tool (see Figure \ref{fig:UI}). The goals were to (1) make citizen-contributed data explorable through visualization, (2) enable users to communicate and share air quality issues with scientific evidence, and (3) empower community members to make evidence-supported decisions. EHC enables exploring and sharing compelling scientific evidence of local environmental impacts of oil and gas drilling activities interactively.

\begin{figure}[t]
	\centering
	\includegraphics[width=1\columnwidth]{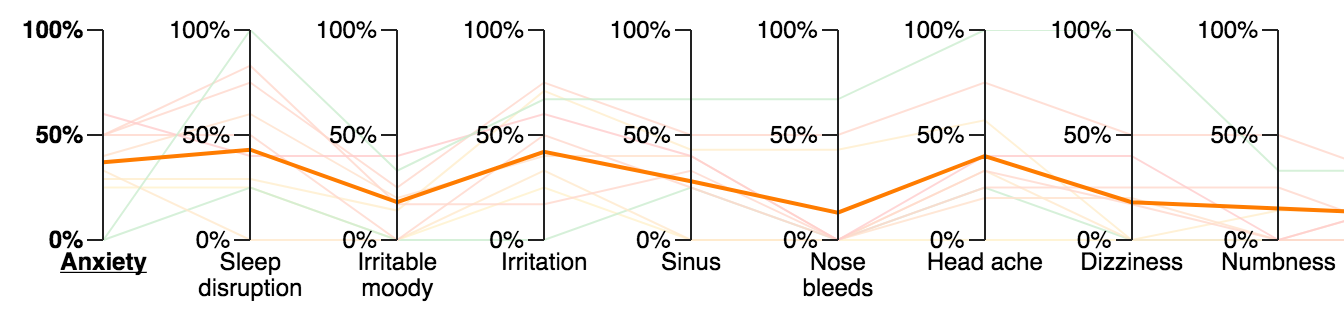}
	\vspace{-6mm}
	\caption{When selecting health data by clicking on the top-left button in Figure \ref{fig:UI}, the bottom parallel coordinate plot changes.}~\label{fig:UI-part-health}
	\vspace{-8mm}
\end{figure}

EHC is an interactive system that supports citizen science, where amateurs and professionals form partnerships through actual participation or collaboration in producing scientific knowledge \cite{Louv-2012, Irwin-1995, Bonney-2014, Bonney-2016, McKinley-2015, EU-2013}. Several existing tools equip citizens with the capabilities to curate data and share data-driven scientific knowledge among stakeholders and community members. Creek Watch \cite{Kim-Robson-2011} is a mobile and web application tool for reporting images and descriptions about the local waterway condition to assist water management policymaking. Sensr \cite{Kim-2013} is a framework for creating mobile applications to collect and manage citizen-contributed data without coding skills. Tian et al. \cite{Tian-2016}, Kim et al. \cite{Kim-2010}, and Kuznetsov et al. \cite{Kuznetsov-2011} implemented air quality monitoring systems that measure particulate matters with commercial or customized sensors and visualize these data in browsers or mobile applications. Our work is distinguished from these projects by two aspects. First, EHC provides scientific evidence from various perspectives by visualizing multiple types of data from air quality sensors, health surveys, and personal stories. Second, instead of showing raw data, EHC aggregates data temporally and geographically to enable comparing different local areas.

\section{System}

During system development, we collaborated with health professionals from the non-profit organization in implementing system features. We began the design process by investigating the data types that the non-profit organization gathered from affected residents, as different data types require distinct visualization affordances. There were three data types: air quality metrics, self-reported health symptoms, and personal stories with images. Since 2014, the non-profit organization has provided portable air quality sensors \cite{Speck, Taylor-2015} to affected residents. After a month of placing sensors indoors and outdoors, the organization collected the sensors, computed air quality statistics from the raw sensor values, and presented these statistics to affected residents in report form. Also, affected residents filled out a self-reporting health survey to indicate physical and psychosocial symptoms that they experienced during the period when sensors were placed. The organization interviewed several affected residents about their personal stories of living near oil and gas drilling sites and collected photographs of their home environments. From these interviews, the organization created a series of photos with narrative text. Integrating the sensor, survey, and interview data into EHC posed privacy issues. To protect the privacy of participants, we de-identified and aggregated data based on zip code boundaries. This approach addressed the concern that confidentiality could be compromised by re-identification of data. EHC stored these de-identified data in a Google Sheet, which enabled the stakeholders to work collaboratively on adding more citizen-contributed data in the future with ease without programming skills. To automate the process of updating data, a Python script on the server periodically parsed the Google Sheet data into suitable formats for each visualization.

\begin{figure}[t]
	\centering
	\includegraphics[width=1\columnwidth]{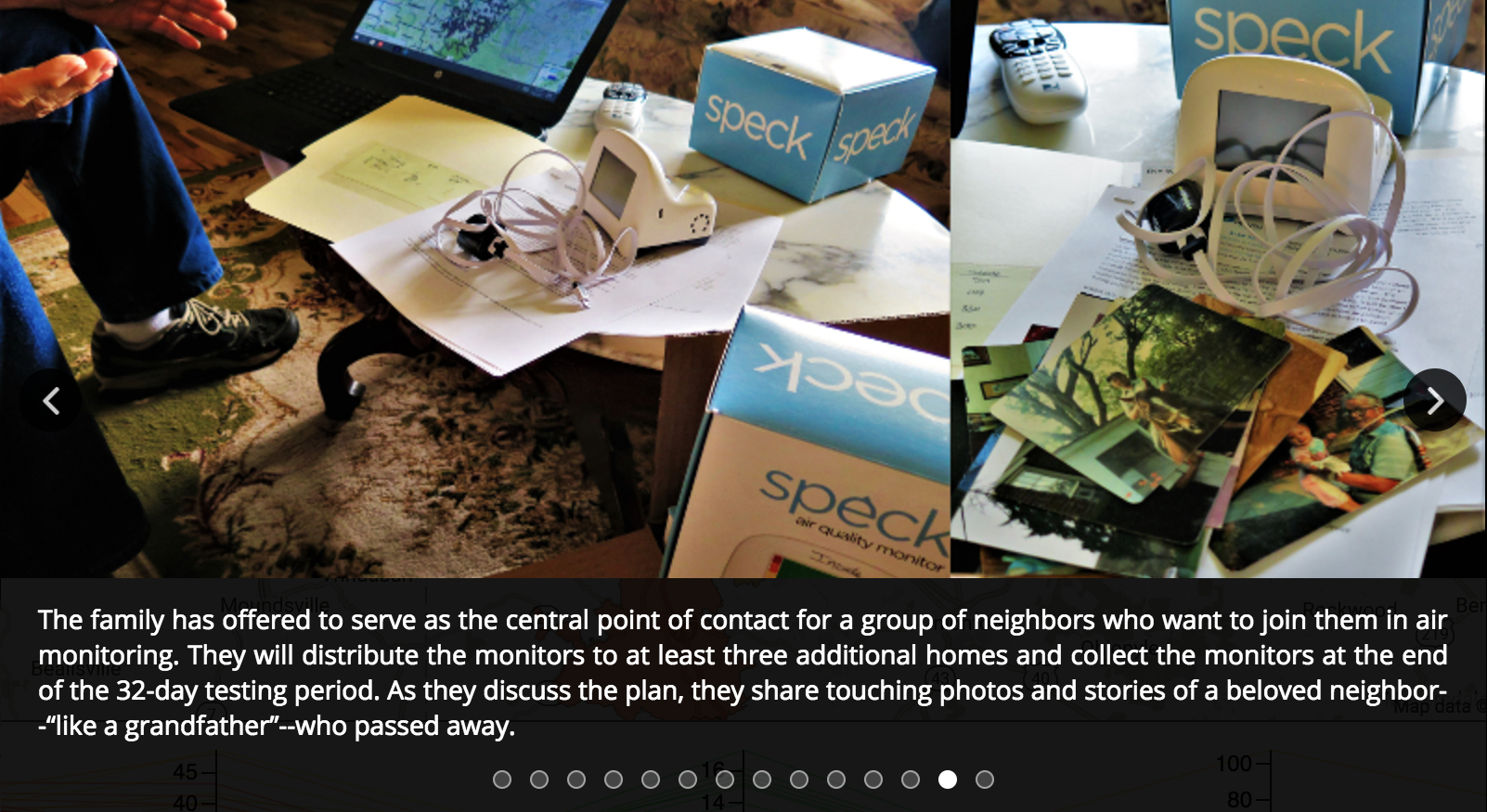}
	\vspace{-6mm}
	\caption{The image slider of personal stories from residents.}~\label{fig:story}
	\vspace{-8mm}
\end{figure}

EHC permits reviewing and comparing aggregated data among different regions simultaneously. To enable interpreting patterns and identifying key policy issues from multiple types of data, we implemented a \textbf{heatmap}, a \textbf{parallel coordinate plot}, and a \textbf{story slider} in HTML/JavaScript. The \textbf{heatmap} (see the top part of Figure \ref{fig:UI}) contains colored polygons to indicate zip code regions which contain air quality sensor data. A color legend (see bottom-right of the map part of Figure \ref{fig:UI}) displays the relative color scale from green, yellow, orange, to red, which corresponds to -1, -0.5, 0.5, and 1 standard deviation away from the mean value respectively. When users click on a colored zipcode, an information window shows up to provide summary statistics of air quality data in the corresponding zip code region. The \textbf{parallel coordinate plot} \cite{Inselberg-1990} (see the bottom part of Figure \ref{fig:UI}) displays the distribution of summary statistics describing air quality or health data. Each axis of the plot represents one statistic, such as the average number of air quality peaks per day. This plot allows users to visually compare relative values of a statistic across different zip code regions. For instance, when the number of peaks per day is selected (see Figure \ref{fig:UI}), red-colored zip code regions on the map have a relatively higher number of peaks per day than all other regions. Users can select a statistic by clicking on the corresponding label on the axis. The \textbf{story slider} (see Figure \ref{fig:story}) shows personal stories and images collected from interviews. This combined visual and narrative presentation offers insight into personal experiences with oil and gas exposures and their involvement with air monitoring. Users can click on open-book icons on the heatmap to explore stories on the slider.

\section{Evaluation}

We conducted a 2-hour focus group study \cite{Langford-2003, Rennekamp-2000, Krueger-2014} and applied affinity diagramming \cite{Lucero-2015, Baxter-2015, Beyer-1997, Holtzblatt-2004, Kawakita-1991} to gain insights about: (1) potential issues about system features and (2) affordances that EHC provided or would support in the future. Seven air quality experts were invited to discuss EHC with a software developer and three health professionals. We found that the discussion was centered around three themes found in previous research \cite{DiSalvo-2008, Kuznetsov-2011, Kuznetsov-2010, Kim-2010, Kim-Robson-2011, Kim-2013}: exploration, investigation, and advocacy. First, \textbf{exploration} refers to supporting the understanding of air quality variables, data sources, and visualizations. For instance, participants mentioned the importance of providing instructions and explanations to users about the provided sensor statistics and the health variables. Participants also suggest that the color red should always indicate a qualitatively worse situation as it relates to potential health impacts, instead of a numerically higher value. Second, \textbf{investigation} pertains to recognizing and comparing data patterns, forming hypotheses, and building narratives with evidence. For example, providing methods for simultaneously comparing health and air quality data is critical for allowing users to investigate the hypotheses that interest them. Additionally, participants recommended adding background variables, such as demographics, to provide more context and enhance scientific evidence. Third, \textbf{advocacy} refers to validating data, taking actions with scientific evidence, and advocating for social impact and political change. For instance, as stories are compelling in evoking emotions and may leave users with the desire to take action, participants suggested adding resources at the end of the story slider to encourage community engagement. Moreover, participants pointed out that there is a need for abstracting data and visuals into concise and convincing reports that can easily be shared with stakeholders and raise the awareness of air quality issues.

\section{Discussion and Future Work}

EHC has been deployed in the local community affected by oil and gas development. Although EHC is being iteratively improved, it enables and encourages health professionals in the non-profit organization to add, visualize, and share incoming data interactively among stakeholders and citizens without the assistance of computer scientists. With the help of air quality experts and health professionals, we have conducted a focus group study to understand issues about system features and determined possible future directions. The result supports the findings in previous research conducted by DiSalvo \cite{DiSalvo-2008}, Kuznetsov \cite{Kuznetsov-2010, Kuznetsov-2011}, and Kim \cite{Kim-2010, Kim-Robson-2011, Kim-2013}. As participants in this study were limited to experts, the result does not reflect the opinions of users with other levels of participation and expertise, such as residents or the general public. Future work will involve conducting more focus group studies to receive feedback from a broader audience. Moreover, we have not evaluated the impact of EHC on experts nor residents. Future research is needed to understand motivations of participation and evaluate attitude changes after using EHC, such as changes in the awareness of air quality problems, confidence in reaching goals, and sense of belonging in a community. We hope that this work will lay a foundation for researchers who develop information technology that provides scientific evidence from multiple perspectives to empowers citizens.

\section{Acknowledgments}

 The Pittsburgh Foundation, the Southwest Pennsylvania Environmental Health Project (Jessa Chabeau, Ken Hamel, Raina Rippel, and Jill Kriesky), and the Global Communication Center of Carnegie Mellon University.

%\balance{} 

\bibliographystyle{SIGCHI-Reference-Format}
\bibliography{sample}

\end{document}